\documentclass[12pt,preprint]{aastex}
\usepackage{natbib}
\usepackage{amssymb,amsmath} % needed for $\mathcal{R}$

\shortauthors{Huang et al.}
\begin{document}
\title{Afterglow Synchrotron Radiations follow the $L_{\rm p, iso}-E_{\rm p,z}-\Gamma_0$ relation of Gamma-Ray Bursts? Cases of GRBs 190114C, 130427A, and 180720B}
\author{Xiao-Li Huang\altaffilmark{1,2}, En-Wei Liang\altaffilmark{2}, Ruo-Yu Liu\altaffilmark{1}, Ji-Gui Cheng\altaffilmark{2}, Xiang-Yu Wang\altaffilmark{1}}
\altaffiltext{1}{School of Astronomy and Space Science, Nanjing University, Nanjing 210023, China}
\altaffiltext{2}{Guangxi Key Laboratory for Relativistic Astrophysics, School of Physics Science and Technology, Guangxi University, Nanning 530004, China; lew@gxu.edu.cn}

\begin{abstract}
Bimodal spectral energy distributions (SEDs) of gamma-ray burst (GRB) afterglow of GRBs 190114C, 130427A and 180720B confirm that they are originated from the synchrotron emission (Syn) and synchrotron self-Compton Scattering process (SSC) of electrons accelerated in the jets. The radiation mechanism and the physics of the observed spectrum-luminosity/energy relations of GRBs remain as open questions. By extracting the Syn component through fitting their early afterglow SEDs with the Syn+SSC model, we find that their luminosity ($L_{\rm syn}$), peak energy ($E_{\rm p,syn,z}$), and the Lorentz factor of the afterglow fireball ($\Gamma_t$) follow the $L_{\rm p, iso}-E_{\rm p,z}-\Gamma_{0}$ relation of prompt gamma-rays, where $L_{\rm p, iso}$ is the isotropic luminosity, $E_{\rm p, z}$ is the peak energy of the $\nu f_\nu$ spectrum in the burst frame, and $\Gamma_0$ is the initial Lorentz factor of the fireball. To examine whether late afterglows is consistent with this relation, we calculate the synchrotron component at late afterglows. It is found that they also follow the same $L_{\rm p, iso}-E_{\rm p,z}-\Gamma_{0}$ relation, albeit they are not consistent with the $L_{\rm p, iso}-E_{\rm p,z}$ relation. Our results may imply that the $L_{\rm p, iso}-E_{\rm p,z}-\Gamma_{0}$ would be an universal feature of synchrotron radiations of electrons accelerated in GRB jets throughout the prompt and afterglow phases among GRBs. Its origin is not fully understood and possible explanations are briefly discussed.
\end{abstract}


\section{Introduction}
Gamma-ray bursts (GRBs) are the most luminous electromagnetic explosions in the Universe. The radiation mechanism of the GRB prompt gamma-ray emission is still an open question. The proposed models include synchrotron radiation of ultra-relativistic electrons accelerated in the internal shocks (ISs) or internal magnetic dissipation regions (e.g. Kumar \& Zhang 2015). The observed spectra of most GRBs in the keV-MeV band can be empirically fitted with a smooth broken power-law function (the so-called Band function, Band et al. 1993), with exception of a few GRBs whose spectrum is dominated by a thermal component (e.g. GRB 090902B, Ryde \& Pe'er 2009) or a superimposed thermal component to the Band function (e.g. Ryde et al. 2010). However, the synchrotron radiation mechanism cannot naturally explain the observed GRB spectra since the derived low-energy index of the band function usually violate the death line of the synchrotron radiation (e.g. Preece et al. 2000). Uhm \& Zhang (2014) suggested that the magnetic field strength of the GRB fireball may continuously decrease with radius and the fast-cooling electrons can have a harder energy spectrum. More recently, Liu et al. (2020) proposed that the Band function may be attributed to the synchrotron emission of the electrons with a bump-shape distribution in its low-energy regime. Alternately, some authors attempted to interpret the entire Band function as emission from a dissipative photosphere (e.g. Giannios 2008).

Some empirical relations between the peak energy ($E_{\rm p, z}$) of the $\nu f_\nu$ spectrum in the burst frame and burst isotropic luminosity/energy have been found, such as the $E_{\rm p, z}-E_{\rm iso}$ relation (Lloyd et al. 2000; Amati et al. 2002), the $E_{\rm p, z}-L_{\rm p, iso}$ relation (Yonetoku et al. 2004), the $E_{\rm p, z}-E_{\rm jet}$ relation (Ghirlanda et al. 2004), the $L_{\rm p, iso}-E_{\rm p, z}-t_{\rm jet}$ relation (Liang \& Zhang 2005). Liang et al. (2010) discovered a relation between the $E_{\rm iso}-\Gamma_0$, where $\Gamma_0$ is the initial Lorentz factor of the GRB fireball. By incorporating these relations, Liang et al. (2015) revealed a tight $L_{\rm p, iso}-E_{\rm p, z}-\Gamma_0$ relation.  However, the origins of these relations are still not well-understood, although many possible explanations are proposed (Lloyd et al. 2000; Zhang \& M\'{e}sz\'{a}ros 2002; Rees \& M\'{e}sz\'{a}ros 2005; Liang et al. 2010, 2015). We should note that these correlations may be suffered great observational biases, such as sample selection, instrumental truncation, and cosmological evolution (Lloyd et al. 2000; Lloyd \& Petrosian 1999; Dainotti et al. 2018; Dainotti \& del Vecchio 2017; Dainotti \& Amati 2018; see also Dainotti 2019 and references therein). 

On the other hand, the radiation physics of the GRB afterglows is well established. The radio to X-ray afterglow emission is generally explained with synchrotron radiation of electrons accelerated by the external shock (e.g. M\'{e}sz\'{a}ros \& Rees. 1997). Recent detection of GeV-TeV afterglow emission in GRB 190114C solidly confirmed that the afterglows are from synchrotron radiations and synchrotron-self-Compton process (SSC; MAGIC Collaboration et al. 2019b). A spectral harding feature, which may be the evidence for the SSC component, was also observed with Fermi/LAT in the afterglows of GRBs 130427A and 180720B (Liu et al. 2013; Ackermann et al. 2014; Duan et al. 2019; Abdalla et al. 2019).

In this paper, we examine whether the afterglow synchrotron radiations satisfy the empirical relations of the prompt gamma-rays in order to reveal possible connection between the prompt gamma-rays and early afterglows. We describe the data of GRBs 130427A, 180720B, and 190114C in \S 2, and present our broadband spectral energy distribution (SED) modeling in \S 3. We examine the agreement of the their synchrotron radiation components to the $L_{\rm p, iso}-E_{\rm p, z}$, $L_{\rm p, iso}-\Gamma_{0}$ and $L_{\rm p, iso}-E_{\rm p, z}-\Gamma_0$ relations in \S 4. Discussion on our results is presented in \S 5. Our conclusions and presented in \S 6. Throughout, a concordance cosmology with parameters $H_0 = 71$ km s$^{-1}$ Mpc $^{-1}$, $\Omega_M=0.30$, and $\Omega_{\Lambda}=0.70$ is adopted.

\section{Data}
GRB 190114C ($z=0.4245\pm0.0005$; Castro-Tirado et al. 2019) was triggered by the {\em Swift}/BAT, {\em Fermi}/GBM, and Konus-Wind. Its observed time-integrated spectrum in 30 keV-20 MeV band is well fitted by the Band function, which yields $\alpha=-0.73\pm0.02$, $\beta=-3.17_{-0.20}^{+0.16}$, and $E_{\rm p}=646\pm16\rm keV$ (Frederiks et al. 2019). Its peak isotropic luminosity is $L_{\rm p, iso}=(1.67\pm0.05)\times10^{53}$ erg/s in the rest-frame 1-10000 keV band. Bright X-ray and optical afterglows were detected (Ajello et al. 2020). Interestingly, its high energy afterglows were also detected by {\em Swift}/BAT, {\em Fermi}/GBM, {\em Femri}/LAT, and the MAGIC telescope (MAGIC Collaboration et al. 2019a). The SED in the X-ray-GeV-sub-TeV band of the early afterglows in some time slices show a clear bimodal feature, which is well explained with the synchrotron radiation and self-Compton scattering process models (MAGIC Collaboration et al. 2019b). We select the SED observed in 68-110 seconds for our analysis since the afterglow is bright and multiple wavelength data are available in this slice. We read its gamma-ray data from Figure 1 of MAGIC Collaboration et al. (2019b). Note that the X-ray data reported by MAGIC Collaboration et al. (2019b) is in the $\sim 1.5-10$ {\rm keV} band. Since the X-ray spectral index is essential for evaluating the spectral regime of the synchrotron radiations, we therefore extract the X-ray spectrum in the 0.3-10 keV and fit it with an absorbed single power-law function. The HI absorption of both our Galaxy and the GRB host galaxy are taken into account. We obtain $\beta_X=-0.69\pm0.12$ and unabsorbed flux $F_X=1.4\times 10^{-7}$ erg cm$^{-2}$ s$^{-1}$ with a c-stat of 674/751. Due to lack of optical observations in the selected time interval, MAGIC Collaboration et al. (2019b) did not show the SED in the optical band. Optical data is also critical to determine the synchrotron radiation peak. We obtain the optical data through linear interpolation with the V band data in the time interval from 58 to 626 seconds post the BAT trigger (MAGIC Collaboration et al. 2019b). Our re-constructed SED in the time slice 68-110 seconds is shown in Figure 1, where the unabsorbed X-ray spectrum is showed as a bow tie.

GRB 180720B ($z=0.654$; Vreeswijk et al. 2018) was triggered by {\em Swift}/BAT (Siegel et al. 2018), {\em Fermi}/GBM (Roberts et al. 2018), and Konus-Wind (Frederiks et al. 2018). Its time-integrated spectrum in 20 keV-15 MeV band can be well fitted by the Band function, which yields $\alpha=-1.01\pm0.06$, $\beta=-2.07_{-0.08}^{+0.07}$, and $E_{\rm p}=451_{-45}^{+52}\rm keV$. Its peak isotropic luminosity is $L_{\rm p, iso}=1.8\times10^{53} \,\rm erg/s$ (Frederiks et al. 2018). The afterglow emission in the keV-MeV-GeV band was simultaneously observed in the time slice of 94-220 seconds. Therefore, we select the SED in this time slice for our analysis. We read the GBM and LAT data in this time slice from Duan et al. (2019). The derived SED is shown in Figure 1. Note that the XRT clearly detected the X-ray afterglow in this time slice, but a bright flare dominates the X-ray emission in this time interval. Therefore, we do not add the XRT data to its SED.

GRB 130427A ($z=0.34$; Maselli et al. 2014) was triggered by {\em Swift}/BAT (Maselli et al. 2013), {\em Fermi}/GBM (Zhu et al. 2013), and Konus-Wind (Golenetskii et al. 2013). Its spectrum observed with Konus-Wind in the 20 keV-15 MeV band can be well fitted by the Band function, which yields $\alpha=-0.958\pm0.006$, $\beta=-4.17\pm0.16$, and $E_{\rm p}$ is $(1028\pm8) \rm keV$ (Golenetskii et al. 2013). Its peak isotropic luminosity is $L_{\rm p,iso}\sim2.7\times10^{53} \rm erg/s$ in the $1-10^4$ keV band. Its afterglows were detected in the optical, X-ray, and GeV bands (Ackermann et al. 2014; Maselli et al. 2014). Its afterglow spectrum is hard in the GeV band, implying an extra component over the emission in the optical-X-ray band, which would be resulted from the SSC process (Fan et al. 2013; Liu et al. 2013; Ackermann et al. 2014).
We select the early afterglow in the time interval 138-750 seconds for our analysis, in which the hard SSC emission component is clear detected.
We read the SED data in the gamma-ray band in this time slice from Ackermann et al. (2014). As mentioned above, the optical and X-ray data are crucial for determining the synchrotron radiation peak, we construct its SED in the optical, X-ray and GeV band in the time slice of 138-750 seconds. We extract the X-ray data observed with {\em Swift}/XRT and fit the spectrum with a single power-law. We obtain $\beta_X=-0.55\pm+0.02$ and unabsorbed flux of $F_X=2.1\times 10^{-8}$ erg cm$^{-2}$ s$^{-1}$ with a c-stat of 1078/801. The optical data are taken from Perley et al. (2014), Laskar et al. (2013) and Vestrand et al. (2014). The derived SED is shown in Figure 1 and the unabsorbed X-ray data are shown with a bow tie.

\section{SED modeling}
\begin{figure}[htbp]
\centering
\includegraphics[width=0.5\textwidth, angle=0]{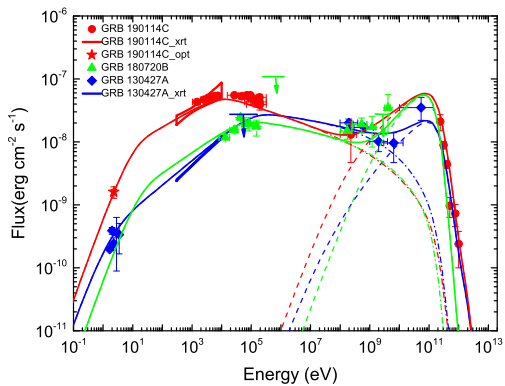}
\caption{Observed broad-band SEDs of GRBs 130427A at $t=138-750$ s (blue dots, the bow tie, and the upper limit), 190114C at $t=68-110$ s (red dots and the bow tie), and 180720B at $t=94-220$ s (green dots and the upper limit) together with our fits using the models of electron synchrotron radiations and the SSC process (lines): {\em solid}---the sum of all emission components, {\em dash-dotted}---the synchrotron component, and {\em dashed}---the SSC component.}
\label{GRBs}
\end{figure}

As illustrated in Figure~\ref{GRBs}, the bimodal feature is clearly seen in the SED of the afterglows of GRB 1901014C and marginally detected in GRBs 130427A and 180720B (see also Liu et al. 2013; Duan 2019). We model the  SEDs at some selected time intervals with the synchrotron radiation and SSC models for electrons accelerated via external forward shocks (Sari \& Esin 2001; MAGIC Collaboration et al. 2019b and Liu et al. 2013). The Klein-Nishina effect is taken into account (Nakar et al. 2009; Wang et al. 2010). Extragalactic background light (EBL) absorption model is taken as that reported by Franceschini et al. (2008). The surrounding medium is assumed to be interstellar medium (ISM) with density $n$. The distribution of the radiating electrons is taken as a single power-law function, i.e., $\mathrm{d}N/\mathrm{d}\gamma_{e}\propto\gamma_{e}^{-p}$, where $\gamma_e$ is the Lorentz factor of the electrons. The initial Lorentz factor of the GRB fireball ($\Gamma_0$) is estimated with the afterglow onset peak time (Ravasio et al. 2019; Ackermann et al. 2014).

Our SED fits are shown in Figure~\ref{GRBs}. One can find that the SEDs are well fitted with the model. For GRBs 130427A and 180720B, their SSC components are only detectable with {\em Fermi}/LAT. Due to the EBL absorption, the observed SSC peaks are $\sim 100$ GeV. Although their SSC components are quite similar, the synchrotron peak of GRB 190114C is brighter than that of GRBs 130427A and 180720B. Their SEDs in the sub-GeV band are dominated by the synchrotron radiations. The optical emission is located in the synchrotron self-absorbtion regime. The X-ray emission and soft gamma-rays are attributed to the synchrotron component, peaking at around tens to hundreds keV. The derived model parameters, including isotropic kinetic energy ($E_{\rm k, iso}$), the energy partition factors of the electrons ($\epsilon_e$) and the magnetic field $\epsilon_B$, the Lorentz factor of the fireball at a selected time interval ($\Gamma_t$), $\Gamma_0$, $n$, and $p$ are reported in Table~\ref{SED_modeling}. The model parameters are quite similar among the three GRBs, i.e., isotropic kinetic energy $E_{\rm k, iso}=(0.65- 2.5)\times 10^{54}$ erg, $\epsilon_e=0.02- 0.1$, $\epsilon_B=(0.4- 2)\times 10^{-4}$, $\Gamma_0=448-540$, $n=0.4- 1$ cm$^{-3}$, $p=2.22- 2.45$.

\section{Correlations among the Luminosity, Peak energy and Lorentz Factor for the Prompt Gamma-Rays and Afterglow Synchrotron Radiations}
The physics of the empirical relations among $L_{\rm p, iso}$, $E_{\rm p,z}$, and $\Gamma_0$ of GRBs is still unclear. In this section, we examine whether the extreme bright GRBs 130427A, 180720B, and 190114C are consistent with this relation. Furthermore, it was speculated that the dominated radiation mechanism of GRBs would be synchrotron radiations of relativistic electrons accelerated in their jets. Therefore, we also test if the synchrotron emission component in the afterglows of GRBs 130427A, 180720B, and 190114C agrees with these empirical relations. We make regression analysis for deriving these correlations with the Bayesian method and the Markov Chain Monte Carlo (MCMC) numerical technique (e.g. D'Agostini 2005). If the upper and lower uncertainties are asymmetrical, we adopt the error as their average. In case of that they are not available, we adopt a relative error of 20\% in our analysis. 

Based on our SED modeling results, we measure the peak energy ($E_{\rm p, syn}$) and isotropic luminosity ($L_{\rm syn}$) in the $1-10^4$ keV band of the synchrotron components at the selected time interval. We obtain $E_{\rm p, syn}=170,\ 88,\ {\rm and}\ 14$ keV, and $L_{\rm syn}=2.7\times 10^{50}, \ 4.1\times 10^{50}$, and $3.0\times 10^{50}$ erg/s for GRBs 130427A, 180720B, and 190114C, respectively. The Lorentz factor of the fireball at the given time is calculated with
$\Gamma_t=31.2\,E_{\rm k,52}^{1/8}\,n^{-1/8}\,t_{3}^{-3/8}$,
where the notation $Q_n$ is defined as $Q/10^{n}$ in the cgs unit. Since $\Gamma_t$ is a function of $t$, the $\Gamma_t$ value of a selected time interval $[t_A, t_B]$ is calculated at the middle of the time interval and the uncertainty is taken as its deviations to the $\Gamma$ values at $t_A$ and $t_B$. Therefore, we get $\Gamma_t=62^{+35}_{-12}, \ 93^{+19}_{-13}, \ {\rm and}\, 146^{+18}_{-12}$ for GRBs 130427A, 180720B, and 190114C in the selected time intervals.

Figure 2 shows the prompt gamma-rays and the synchrotron radiations of the afterglows of the three GRBs in the $L_{\rm p, iso}-E_{\rm p, z}$ and $L_{\rm p, iso}-\Gamma_0$ planes together with a GRB sample from Liang et al. (2015), where $L_{\rm p, iso}$ is taken as $L_{\rm syn}$, and $E_{\rm p, z}$ is taken as $E_{\rm syn, z}$, and $\Gamma_0$ is taken as $\Gamma_{t}$ at the selected time interval for the afterglows. Combining the data of the three GRBs and the GRB sample from Liang et al. (2015), we obtain $L_{\rm p, iso}\propto E^{2.20\pm 0.20}_{\rm p,z}$ and $L_{\rm p, iso}\propto \Gamma_0 ^{2.56\pm 0.23}$. One can observe that both the prompt gamma-rays and the afterglows synchrotron radiations of the three GRBs follow the $L_{\rm p, iso}-E_{\rm p, z}$ and $L_{\rm p, iso}-\Gamma_0$ relations within the $3\sigma$ dispersion region.

Liang et al. (2015) discovered a tight relation among $L_{\rm p, iso}$, $E_{\rm p,z}$ and $\Gamma_{0}$ for the GRB prompt gamma-ray emission. With the data from Liang et al. (2015), we make the regression analysis by employing the Bayesian method and the MCMC techniques (D'Agostini 2005), which yields $L^{r}_{\rm p, iso, 52}=10^{-6.46\pm0.56}\times E^{1.37\pm 0.24}_{\rm p, z} \times \Gamma_{0}^{1.30\pm0.31}$, where the errors are in $1\sigma$ confidence level. Adding the prompt gamma-ray data of the three GRBs to the sample, our analysis gives
\begin{equation}
L^{r}_{\rm p, iso, 52}=10^{-6.50\pm0.53}\times E^{1.37\pm 0.23}_{\rm p, z} \times \Gamma_{0}^{1.33\pm0.31}.
\end{equation}
One can observe that the parameters of the new $L_{\rm p, iso}-E_{p,z}-\Gamma_0$ relation are consistent with that derived from the GRB sample of Liang et al. (2015) within $1\sigma$ error. Figure 1 shows the linear relation between $L^{r}_{\rm p, iso, 52}$ and  $L_{\rm p, iso, 52}$ and its $3\sigma$ dispersion derived from our analysis.

We calculate $L^{r}_{\rm p,\ syn}$ of early synchrotron afterglows at selected time intervals of the three GRBs with Eq. (1) by replacing $E_{\rm p,z}$ and $\Gamma_0$ with $E_{\rm syn, z}$ and $\Gamma_t$, respectively. Adding them to Figure 3, one can find that they are consistent with the $L_{\rm p, iso}-E_{p,z}-\Gamma_0$ relation within $3\sigma$ dispersion region.
Including the early synchrotron afterglow data of the three GRBs, we make regression analysis for deriving the $L_{\rm p, iso}-E_{p,z}-\Gamma_0$ relation again. We have $L^{r}_{\rm p, iso, 52}=10^{-6.66\pm 0.54}\times E^{1.31\pm 0.20}_{\rm p, z} \times \Gamma_{0}^{1.46\pm0.27}$. It is found that the derived parameters also agree with that of Eq. (1) within $1\sigma$ confidence level. 

These results hint that this relation may be universal in both the prompt and very early afterglow phases. We examine whether the synchrotron emission component in late afterglows still follows this relation by calculating its $L_{\rm p, syn}$, $E_{\rm syn, z}$, and $\Gamma_t$ for some selected time intervals based our SED fitting results. The selection of the time slices at the very early stage for our analysis depends on the simultaneous multiple wavelength observations for creating the observed broad band SEDs, as mentioned in \S 2. These time slices are 48-77, 58-133, 102-572 seconds post the GRB trigger in the rest frame for GRBs 190114C, 180720B, and 130427A, respectively. For late afterglows, we select common time slices for the three GRBs, i.e., $(5- 5.1)\times10^{3}$ s, $(5- 5.1)\times10^{4}$ s, $(5- 5.1)\times 10^{5}$ s in the burst rest frame. We derived $L^{\rm r}_{\rm p,syn}$ in these time slices with Eq. (1). The data of these late afterglows are also shown in Figures 2 and 3 with open squares.
It is found that GRBs 130427A and 180720B are significantly out of the 3$\sigma$ region of the $L_{\rm p, syn}-E_{\rm p}$, but they are within the $3\sigma$ dispersion region of the $L_{\rm p, iso}-\Gamma_0$ relation. On the contrary, GRB 190114C is marginally out of the $3\sigma$ region of the $L_{\rm p, iso}-\Gamma_0$ relations, but is within the 3$\sigma$ region of the $L_{\rm p, syn}-E_{\rm p}$. Interestingly, the synchrotron components of the late afterglows of the three GRBs are within the $3\sigma$ dispersion of Eq. (1).

\section{Discussion}
Our above results suggest that the synchrotron radiations of early afterglows may follow the same $L_{\rm p, iso}-E_{\rm p,z}-\Gamma_0$ relation found for the prompt gamma-ray emissions. Note that the prompt emission arises from the internal dissipation, while the afterglow emission arise from the expanding blast waves interacting with the surrounding medium. Thus, the $L_{\rm p, iso}-E_{\rm p,z}-\Gamma$ relation may reflect some connection between the internal shocks and external shocks of the jets.

For internal shocks the energy that goes into the newly shocked electrons per unit observer time is
\begin{equation}
L_e=\epsilon_e L_{\rm sh}=\epsilon_e 4 \pi R^2 c U'_{\rm sh} \Gamma^2
\end{equation}
where $U'_{\rm sh}$ is the energy density of the shock in the comoving frame, $R$ is the radius of the internal shocks, $\Gamma$ is the Lorentz factor of the initial ejecta, and $\epsilon_e$ is the electron equipartition factor. The synchrotron radiation luminosity is $L_{\rm syn}=\eta L_e$, where $\eta$ is the radiation efficiency.

The peak energy of the synchrotron emission is
\begin{equation}
E_{\rm p, syn}=h\nu_p=\Gamma \gamma_p^2 \frac{h e B}{2\pi m_e c},
\end{equation}
where $\gamma_p$ is the Lorentz factors of the electrons with the synchrotron radiation peaking at $E_{\rm p, syn}$, $h$ is the Plank constant, $e$ is the electron charge, $m_e$ is the electron mass, and $c$ is the speed of light. The magnetic field in the internal shock or in the forward shock is given by
\begin{equation}
B=(8\pi \epsilon_B U'_{\rm sh})^{1/2}.
\end{equation}

Based on these equations, we get
\begin{equation}
L_{\rm syn, 52}=\mathcal{A}E_{\rm p, syn, 2}\,\Gamma_2,
\end{equation}
where $\mathcal{A}\equiv 10^{-0.40} \eta \left(\frac{\epsilon_{e,-1}}{\epsilon_{B, -3}}\right) \left(\frac{B_3 R_{14}^2}{\gamma_{p, 3}^2}\right)$. One can find that the power-law indices of $E_{\rm p, syn}$ and $\Gamma$ in Eq. (5) roughly agree with the power-law indices of $E_{\rm p, z}$ and $\Gamma_0$ in Eq. (1). The tight empirical relation may suggest that $\mathcal{A}$ would be quasi-universal during the prompt and afterglow phases among GRBs.

The fact that both the early afterglow and late afterglows satisfy the $L_{\rm syn}-E_{\rm p, syn}-\Gamma_t$ relation  may  place constraints on physical parameters in the GRB shock. In the standard GRB afterglow models, the $\nu F_\nu$ synchrotron radiation spectrum peaks at $\nu_m$ in the fast cooling scenario, or at $\nu_c$ in the slow cooling scenario.  We find that the afterglows of the three GRBs are in the slow cooling scenario, so the flux density
is given by
\begin{equation}
\left\{
\begin{array}{ll}
F_{\nu}=(\nu/\nu_{m})^{-(p-1)/2}F_{\nu,\rm max},\quad\;\qquad \nu_m<\nu<\nu_{c},\\
F_{\nu}=(\nu_{c}/\nu_{m})^{-(p-1)/2}(\nu/\nu_{c})^{-p/2}F_{\nu,\rm max},\quad \nu>\nu_{c},
\end{array}
\right.
\end{equation}
where
\begin{equation}
\begin{array}{ll}
F_{\nu,\rm max}=1.6\times10^{-26}(1+z)D_{28}^{-2} \times \epsilon_{B,-2}^{1/2}E_{\rm k,52}n^{1/2}\,\rm erg\, cm^{-2}s^{-1}Hz^{-1},
 \end{array}
\end{equation}
and the two characteristic frequencies ($\nu_m$ and $\nu_c$) are given by
\begin{equation}
\nu_m=2.6\times10^{15}\,(\frac{p-2}{p-1})^{2}(1+z)^{1/2}E_{\rm k,52}^{1/2}\epsilon_{B,-2}^{1/2}\epsilon_{e,-1}^{2}t_{3}^{-3/2}\,\rm Hz
\end{equation}
and
\begin{equation}
\nu_c=5.6\times10^{16}(1+Y)^{-2}(1+z)^{-1/2}\epsilon_{B,-2}^{-3/2}E_{\rm k,52}^{-1/2}n^{-1}t_{3}^{-1/2}\,\rm Hz.
\end{equation}

In this scenario, we have $E_{\rm p, syn}=h\nu_c\propto(1+Y)^{-2}t^{-1/2}$,  $L_{\rm syn}\propto \nu_c F_{\nu}(\nu_c)\propto (1+Y)^{p-3}t^{-p/2}$, where $Y$ is the Compton parameter for electrons radiating the photons with frequency of $\nu_c$. The Lorentz factor evolves as $\Gamma_t\propto t^{-3/8}$ if the fireball expands adiabatically. Basing on the Eq. (1), we have  $L^{\rm r}_{\rm syn, p}\propto E_{\rm p}^{1.37}\Gamma_t^{1.33}\propto (1+Y)^{-2.74}t^{-1.18}$. Thus, one can infer that $(1+Y)^{p-3}t^{-p/2}\sim (1+Y)^{-2.74}t^{-1.18}$, hence $(1+Y)^{p-0.26}t^{-p/2+1.18}={\rm constant}$. This condition is generally satisfied, given that $p\sim 2.36$ and $Y<1$ for the parameter used in the modeling of the three GRBs (Liu et al. 2013, Wang et al. 2019).

\section{Conclusions}
Using the prompt gamma-ray observational data of GRBs 130427A, 180720B, and 190114C observed with Fermi/GBM and Konus-Wind, we have illustrated that the prompt gamma-ray emission of the three GRBs follows the $L_{\rm p, iso}-E_{\rm p,z}-\Gamma_0$ relation derived from a sample of typical GRBs. The early afterglow spectral energy distributions (SEDs) of the three GRBs in the optical-X-ray-GeV/sub-TeV band are confirmed to be originated from synchrotron radiations and self-Compton Scattering process of ultra-relativistic electrons accelerated in their jets. We fit the broadband SEDs with the Syn+SSC model to derive their electron synchrotron radiation components. We show that their luminosity and peak energy of the synchrotron component as well as the Lorentz factor of the early afterglows also satisfy this relation. To examining whether the synchrotron radiation component in late afterglows is still consistent with this relation, we calculate the spectral parameters of the late afterglows at some selected time intervals and find that they also follow the same $L_{\rm p, iso}-E_{\rm p,z}-\Gamma_0$ relation. Our results may imply that the $L_{\rm p, iso}-E_{\rm p,z}-\Gamma_0$ would be an universal feature of synchrotron radiations of electrons accelerated in GRB jets throughout the prompt and afterglow phases among GRBs.  Its origin is not fully understood and possible explanations were briefly discussed.

\section*{Acknowledgement}
We thank Xue Feng Wu, Jin Jun Geng for helpful discussion. This work is supported by the National Natural Science Foundation of China (Grant No.11533003, 11851304, and U1731239), by the Guangxi Science Foundation and special funding for Guangxi distinguished professors (2017AD22006), by the NSFC under the Grants No. 11625312 and No. 11851304, by the National Key R\&D program of China under the Grant No. 2018YFA0404203.

\clearpage
\begin{table}
\begin{center}
\tiny
  	\caption{Properties of the GRB jets derived from our SED modeling}
	\label{SED_modeling}
	\begin{tabular}{ccccccccccccccc}
    \hline\hline
    GRB &$z$ & $L_{\rm p, iso}$& $E_p$ & $\Gamma_0$ & $\epsilon_e$ & $\epsilon_B$ & $E_{\rm k, iso}$ &  $n$ &$p$ & interval   & $L_{\rm syn}$&$E_{\rm p, syn}$& $\Gamma_{t}$ \\
   &  & (erg s$^{-1}$) &(keV) & &  &($10^{-4}$) & (erg) & (cm$^{-3}$) &  & (s) & (erg s$^{-1}$) & (keV) &  \\
\hline
 130427A& $0.34$  & $2.7E53$ & $1028\pm 8$&  540 & 0.1 & $0.5$ & $6.5E53$  &   0.5   & 2.22 & 137-750 & $2.7E50$& 170& $62^{+35}_{-12}$ \\
\hline
 180720B& $0.654$& $1.8E53$ & $451^{+52}_{-45}$&  448 & 0.1 & $0.4$ & $1.0E54$  &   1.0  & 2.35 & 94-220 & $4.1E50$& 88& $93^{+19}_{-13}$ \\
\hline
190114C & $0.4245$  & $1.7E53$ & $646\pm 16$&  500 & 0.02 & $2$ & $2.5E54$  &   0.4   & 2.45  & 68-110 & $3.0E50$& 14& $146^{+18}_{-12}$ \\
 \hline
	\end{tabular}
\end{center}
\end{table}

\clearpage

\begin{figure}[htbp]
\centering
\includegraphics[width=0.45\textwidth, angle=0]{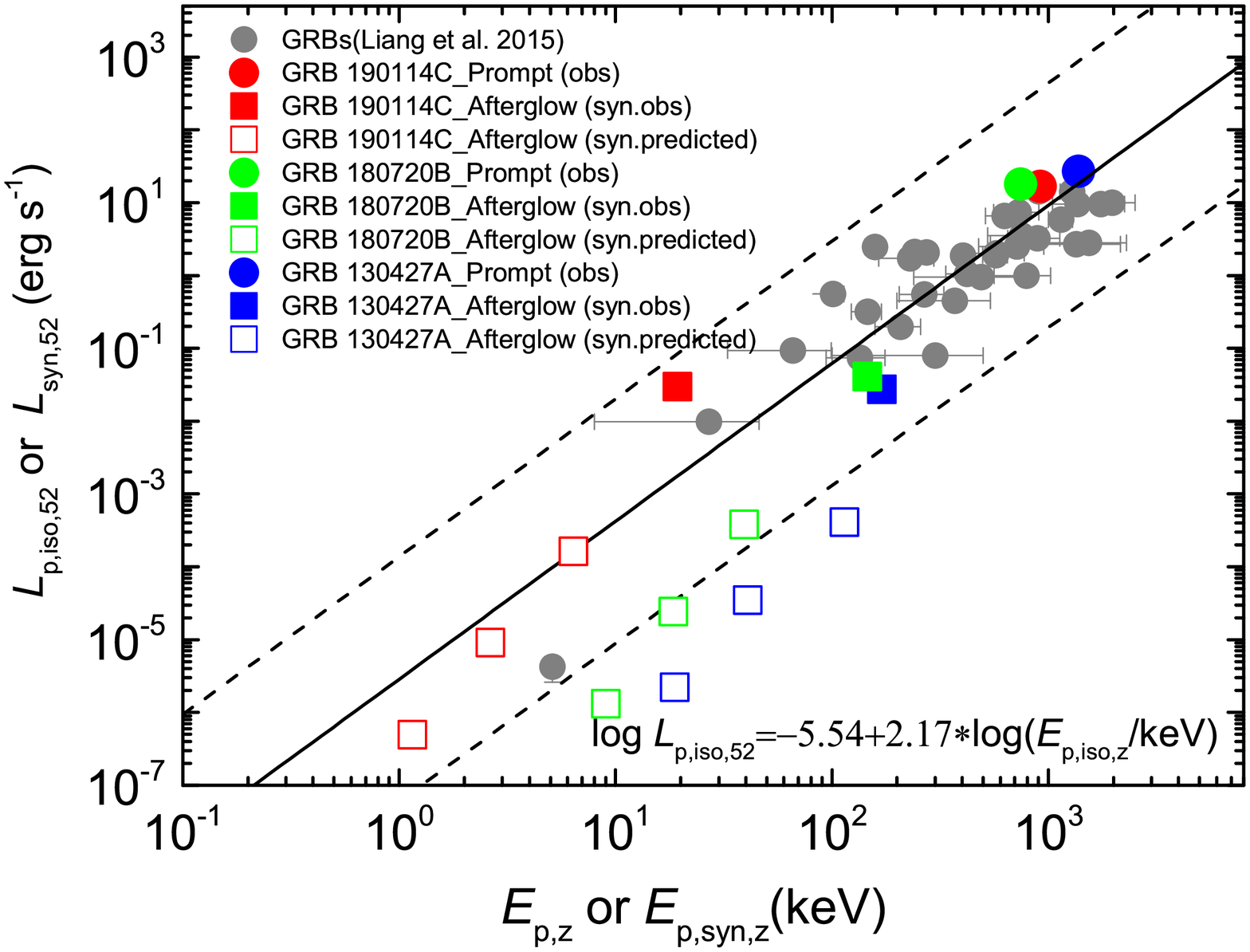}
\includegraphics[width=0.45\textwidth, angle=0]{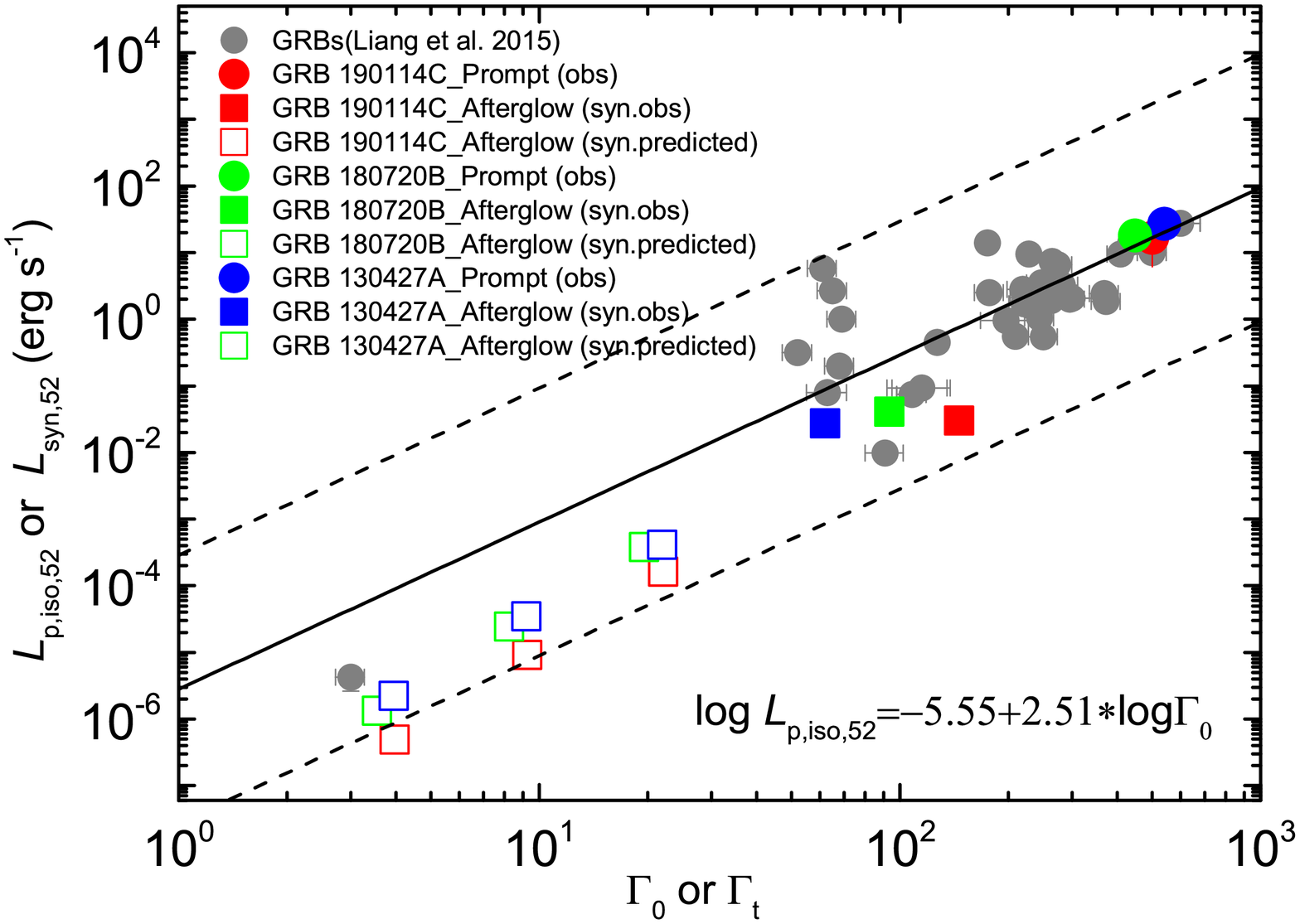}
\caption{$L_{\rm p,iso}$ (or $L_{\rm syn}$) as a function of $E_{\rm p,z}$ (or $E_{\rm p,syn,z}$ ) ({\em left panel}) and $L_{\rm p,iso}$ (or $L_{\rm syn}$) as a function of $\Gamma_{0}$ (or $\Gamma_t$ ) ({\em right panel}) for examining the consistency of the prompt gamma-ray emission (or synchrotron afterglow emission in selected time intervals) of GRBs of 190114C, 180720B, and 130427A to the $L_{\rm p,iso}-E_{\rm p,z}$ and $L_{\rm p,iso}-\Gamma_{0}$ relations derived from our fit to the prompt gamma-ray data ({\em dots}) using the Bayesian method and the MCMC numerical technique (e.g. D'Agostini 2005). The solid and dashed lines are the best fit and its dispersion in 3 $\sigma$ confidence level for the relations.}
\label{Lp-Ep-Gamma}
\end{figure}

\begin{figure}[htbp]
\centering
\includegraphics[width=0.45\textwidth, angle=0]{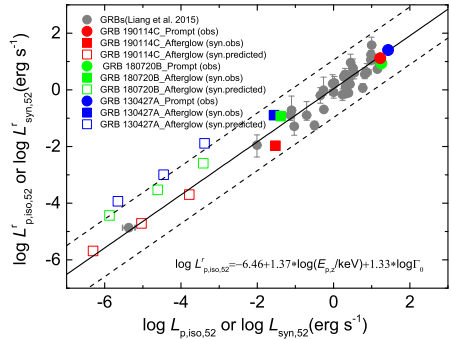}
\caption{
$L^{r}_{\rm p, iso}$ (or $L^{r}_{\rm syn}$) as a function of $L_{\rm p, iso}$ (or $L_{\rm syn}$) for examining the consistency of the prompt gamma-rays and the synchrotron afterglows of GRBs 190114C, 180720B, and 130427A to the $L_{\rm p, iso}-E_{\rm p, z}-\Gamma_{0}$ relation, where $L^{r}_{\rm p, iso}$ (or $L^{r}_{\rm syn}$) is calculated with the  $L_{\rm p, iso}-E_{\rm p, z}-\Gamma_{0}$ relation derived from our fit to the prompt gamma-ray data (dots) using the Bayesian method and the MCMC numerical technique (e.g. D'Agostini 2005). Synchrotron radiation components ({\em open squares}) of the afterglows in the selected time slices of the three GRBs are marked with open squares by taking $L_{\rm p,iso}$ and $E_{\rm p,z}$ as $L_{\rm syn,iso}$ and $E_{\rm syn, p}$, respectively. The {\em solid} and {\em dashed} lines are the best fit and its dispersion in 3 $\sigma$ confidence level for the relation between $L^{r}_{\rm p, iso}$ and $L_{\rm p, iso}$.}
\label{Lp-Ep-Gamma}
\end{figure}

\end{document}